\documentclass[aps,amsmath,amssymb,twocolumn,showpacs]{revtex4}
\usepackage{graphicx,color}
\graphicspath{{./}{./figures/}}
\usepackage{dcolumn}
\usepackage{bm}
\usepackage{array}
\usepackage{hyperref}

\newcommand{\be}{\begin{equation}}
\newcommand{\ee}{\end{equation}}
\newcommand{\bea}{\begin{eqnarray}}
\newcommand{\eea}{\end{eqnarray}}

\newcommand{\ldblbracket}{[\![}
\newcommand{\rdblbracket}{]\!]}

\begin{document}
\bibliographystyle{unsrt}
\title{Revisiting Directed Polymers with heavy-tailed disorder}
\author{Thomas Gueudre}
\affiliation{CNRS-Laboratoire
de Physique Th{\'e}orique de l'Ecole Normale Sup{\'e}rieure, 24 rue
Lhomond,75231 Cedex 05, Paris, France}
\author{Jean-Philippe Bouchaud}
\affiliation{CFM, 21 rue de l'universit\'e, 75007 Paris France}
\author{Pierre Le Doussal}
\affiliation{CNRS-Laboratoire
de Physique Th{\'e}orique de l'Ecole Normale Sup{\'e}rieure, 24 rue
Lhomond,75231 Cedex 05, Paris, France}
\author{Alberto Rosso}
\affiliation{Paris-Sud, LPTMS, UMR 8626, Orsay F-91405, France}
\date{\today\ -- \jobname}


\begin{abstract}
In this mostly numerical study, we revisit the statistical properties of the ground state of a directed polymer in a $d=1+1$ ``hilly'' disorder landscape, i.e. when the quenched disorder has power-law tails. When disorder is Gaussian, the polymer minimizes its total energy through a collective optimization, where the energy of each visited site only weakly contributes to the total. Conversely, a hilly landscape forces the polymer to distort and explore a larger portion of space to reach some particularly deep energy sites. As soon as the fifth moment of the disorder diverges, this mechanism radically changes the standard ``KPZ'' scaling
behaviour of the directed polymer, and new exponents prevail. After confirming again that the Flory argument accurately predicts these exponent in the tail-dominated phase, we investigate several other statistical features of the ground state that shed light on this unusual transition and on the accuracy of the Flory argument. We underline the theoretical challenge posed by this situation, which paradoxically becomes even more acute above the upper critical dimension. 
\end{abstract}
\maketitle

\section{Introduction}

The so-called ``Directed Polymer'' (DP) problem has attracted an enormous amount of interest in the last thirty years \cite{HH_Zhang_review,Meakin1993189,krug_kriecherbauer}. It is stylized model for the pinning of
directed one-dimensional elastic objects (polymers, vortex lines, dislocations,...) by random impurities. It is perhaps the simplest model that captures
the notion of ``frustration'' that is so crucial in many more complex disordered materials, such as spin-glasses: the elasticity of the polymer competes with
the energy of very favorable, but distant pinning sites that would lead to a costly distortion of the polymer. The huge amount of work on this problem is justified
not only because of its intrinsic interest, but also because it can be mapped to a host of other problems: the Stochastic Heat Equation \cite{corwin2012kardar}, itself mapped onto the
Kardar-Parisi-Zhang equation \cite{kardar_dynamic_1986} and the stochastic Burgers' equation
\cite{Bec20071}, population dynamics \cite{PhysRevLett.112.050602}, problems of jammed transport (TASEP)
and crystal growth \cite{krug1991kinetic}.

In 1+1 dimensions (one transverse, one longitudinal, often taken as the ``time'' dimension), the problem is considered to be exactly solved,
at least in some special limits and for some particular observables \cite{johannson_shape,Sasamoto_Spohn,Spohn200671}. It is now well established that in the limit of ``long'' polymers of length $t \to \infty$,
the transverse excursions $x$ are of order $t^{\zeta}$ with $\zeta=2/3$, i.e. much larger than $\sqrt{t}$ that would correspond to the thermal excursion of the polymer in the absence
of disorder. Moreover the total free energy fluctuations scale with an anomalous exponent 
$\theta=1/3$.  Although these scaling exponents have been known at the level of physical rigour since the 80's (using either replica theory \cite{kardar_scaling_1987},
the exact stationary state of the corresponding KPZ equation \cite{huse_huse_1985}, RG techniques \cite{kardar_dynamic_1986} or Mode-Coupling theory \cite{PhysRevLett.86.3946}), it is fair to say that there is up to now no simple,
heuristic derivation of these exponents that would a) unveil the deep physical origin of these results and
b) allow one to extend these results to other, similar problems, such as the Directed Polymer problem in $d+1$ dimensions, for which the situation is still
quite unclear \cite{PhysRevLett.109.170602,PhysRevE.88.042118,PhysRevE.84.061150,PhysRevE.77.031134}. For example, the existence of an upper critical dimension $d_c$ beyond which $\zeta=1/2$ even in the low temperature, pinned phase, is still highly
debated \cite{PhysRevE.85.050103}. In $d=1$ recent results have shown that the total free-energy of the polymer can be written as $-c t + \xi t^{1/3}$, where $c$ is a non universal constant and $\xi$ is proportional to a random
variable with a Tracy-Widom distribution, identical to the one governing the statistics of the largest eigenvalue of random matrices (GOE or GUE, depending on the
boundary conditions) \cite{amir_corwin_2011,dotsenko_EPL,dotsenko_klumov_betheattempt,calabrese_free-energy_2010,calabreseflat}.

In spite of these numerous exact results, it is fair to say that even the 1+1 directed polymer problem is far from understood. Consider for
example the role of the distribution of the pinning energy on the large scale properties of the polymer. One would naively expect that, as with many other problems,
the existence and finiteness of the second moment of the distribution is enough to ensure that the above scaling results (valid for Gaussian or exponential
disorder) hold asymptotically. Surprisingly, though, this does not seem to be the case. A heuristic, Flory-type argument that dates back from the early 90's \cite{zhang_growth_1990,krug_kinetic_1991,biroli_top_2007}
suggest that as soon as the fifth moment of the distribution diverges, one should leave the realm of the standard DP/KPZ $2/3$ scaling, and enter a new regime, where
the extreme values taken by the pinning potential matter and change the scaling results. In fact, the same Flory argument suggests that the situation becomes
worse and worse as the dimension increases \cite{biroli_top_2007}. In fact, any sub-exponential tail of the potential should play a crucial role at and above $d_c$. [It can indeed be checked explicitely that the Derrida-Spohn
solution of the DP on a tree breaks down as soon as the potential has sub-exponential tails \cite{derridaspohn}.] 

The sensitive dependence of large scale properties on the far-tails of the disorder is certainly unusual and highlights our poor grasp of the standard case.
It also raises many technical questions, for example on the validity
of techniques that have been exploited in the context of Gaussian disorder, such as replica method or the functional RG. It is clear that if confirmed, these
far-tailed induced effects would require new, specific theoretical methods that could, indirectly, shed new light on the DP problem altogether. Before embarking on such
a program, we wanted to revisit the 1+1 DP problem with heavy-tailed disorder, and establish numerically, as convincingly as possible,
the violation of the standard DP/KPZ $2/3$ scaling and the corresponding Tracy-Widom statistics. Our results are, quite remarkably, in perfect agreement with
the naive Flory predictions for the diffusion exponent $\zeta$ and the energy exponent $\theta$, which confirms that the value $\zeta=2/3$ only holds if the
probability density of the pinning energy $V$ decays faster that $1/|V|^6$! We study various statistical properties of the DP in the anomalous regime, and attempt to
define and measure certain quantities that directly validates the main assumption of the Flory argument, namely that the accesible extreme values of the pinning
potential dominate the scaling behaviour. We conclude with several open problems. 

\section{\label{model}The model}

\begin{figure}
\includegraphics[scale=0.35]{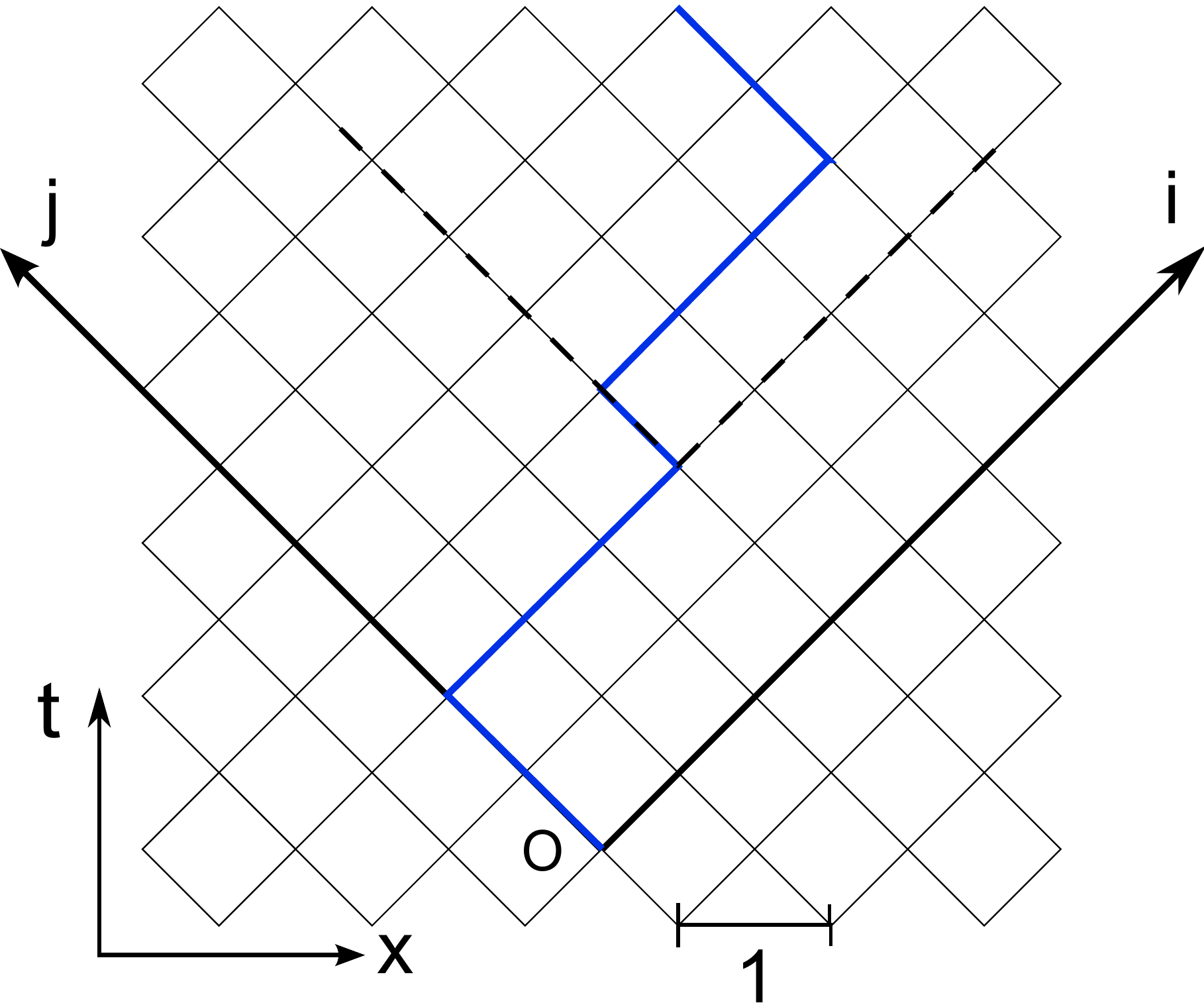}
\centering
\caption{Sketch of the directed polymer model. The blue solid line corresponds to a polymer growing over the square lattice under the hard constraint condition.}
\label{picpoly}
\end{figure}

Here we consider a one dimensional directed polymer growing on the two dimensional square lattice  depicted in Fig. \ref{picpoly}.  

Directed paths grow along the diagonals of the lattice with only $(0,1)$ or $(1,0)$ moves (hard constraint condition), starting in $(0,0)$ and with the second end left free. 
To each site of the lattice is associated a i.i.d. random number $V(x,t)$.
The time coordinate is given by $t=i+j$ and the space coordinate by $x=(i-j)/{2}$.
The total energy of the polymer is the minimum over all paths $\gamma_t$ growing from $(0,0)$ up to time $t$ is defined as
\bea
E(t)=\min _{\gamma_t} \sum _{(x,\tau) \in \gamma_t} V(x,\tau)
\eea
with $2 x \in \ldblbracket -t,t \rdblbracket $ and $ \tau \in \ldblbracket 0,t \rdblbracket $. 
The energy of the polymer satisfies the following transfer matrix recurrence relation:
\bea
E_{x,t+1}=\min (E_{x-\frac{1}{2},t}, E_{x+\frac{1}{2},t}) + V(x,t+1)
\eea
with $E_{x,0}=\delta_{x,0}$. The free end ground state is computed by taking the minimum of the energies over all endpoints $E(t)=\min _x E(x,t)$.  
In this paper, we study the properties of the DP for different disorder distributions $P(V)$, in particular we focus on heavy-tailed pdf decaying as:
\begin{align}
\label{paretotail}
P(V) \sim_{V \to -\infty} \frac{1}{|V|^{1+\mu}}
\end{align}

\begin{figure}
\includegraphics[scale=0.31]{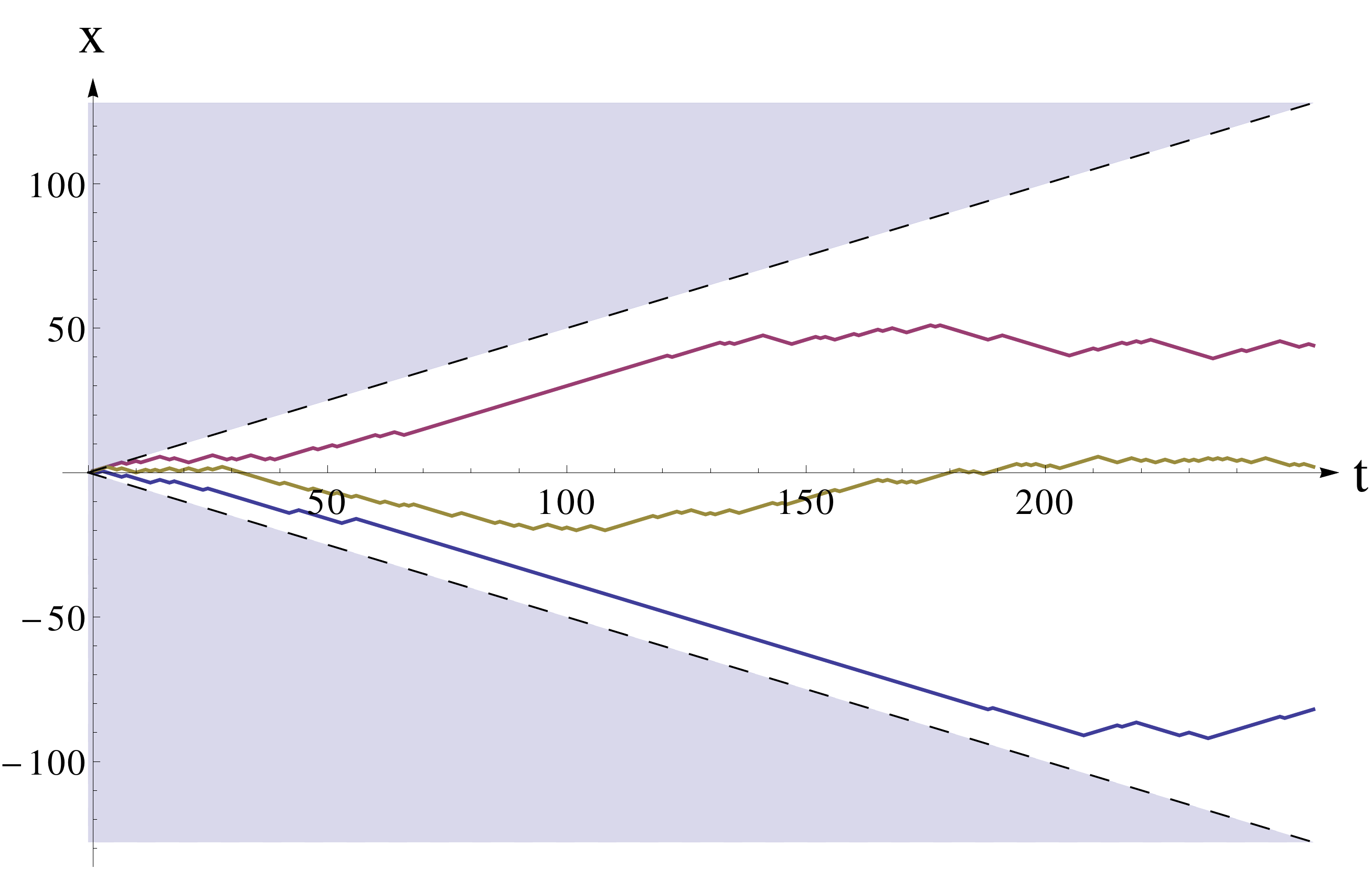}
\centering
\caption{Optimal path ($t=512$) for three different random environments: the Gaussian disorder (in yellow), the heavy-tailed disorder with $\mu = 0.1$ (in blue) and with $\mu = 2.5$ (in red). The shape of the path is strongly affected by the underlying disorder. Because of the hard constrain, the path can only evolve inside the cone delimited by the dashed lines.}
\label{examplespoly}
\end{figure}

\begin{figure}
\includegraphics[scale=0.23]{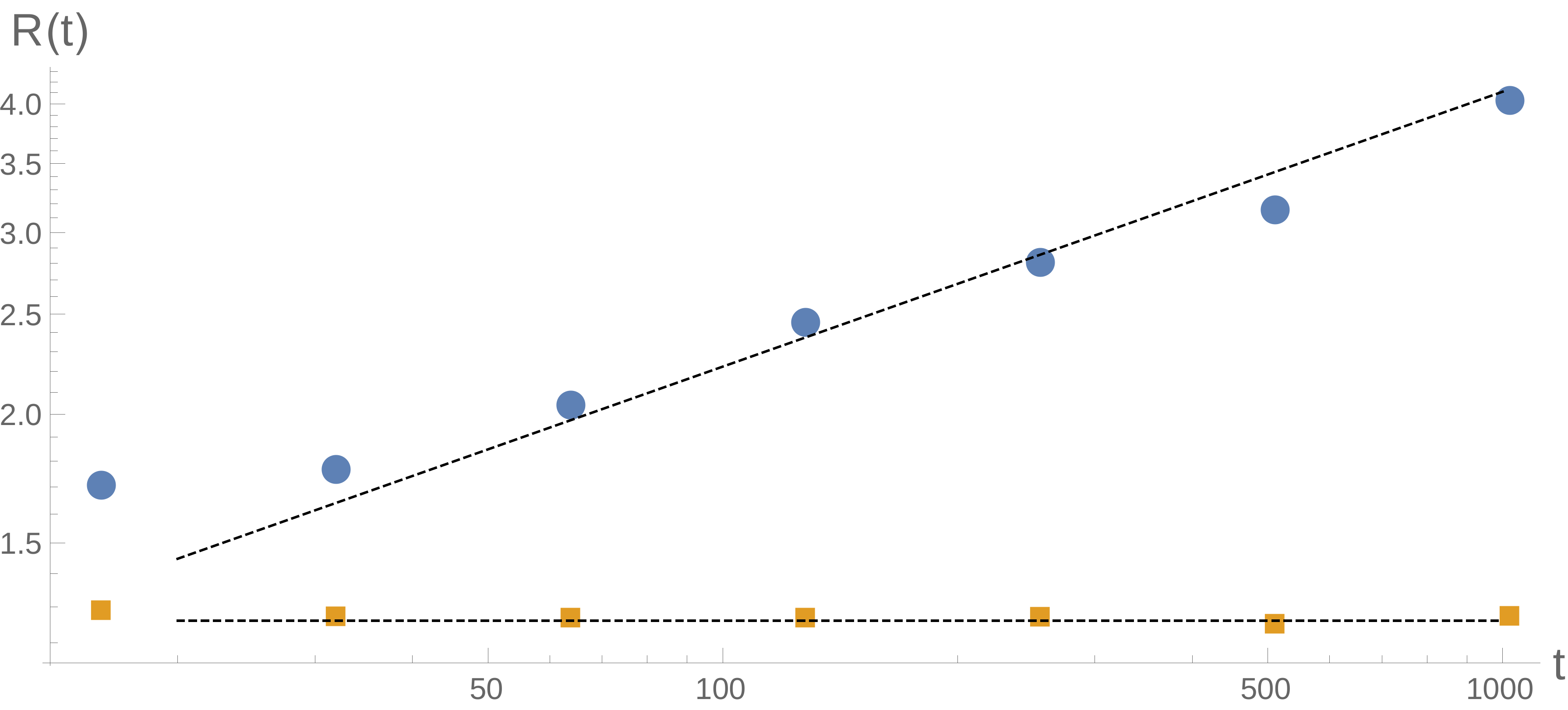}
\centering
\caption{Ratio $R(t)= \min_{\tau < t} V(\tau)/\min_{\tau <t}X(\tau)$ of the minimum energy contribution along the ground state of a polymer of length $t$ to the minimum of a sequence $\lbrace X(\tau) \rbrace$ of $t$ random variables independently drawn from the very same disorder distribution, Eq. (\ref{paretotail}). Blue circles correspond to $\mu = 3$, yellow squares $\mu =8$. In dashed lines are plotted either a constant (for $\mu=8$) or $ t^{\theta_\mu-1/\mu}=t^{4/15}$ (using $\min_{\tau <t}X(\tau) \sim t^{1/\mu}$ and $\theta_3=3/5$. One clearly sees that the optimal polymer for $\mu=8$ does hardly better than for purely random sequences, the $\textit{elitist}$ optimization in $\mu=3$ leads to power-law growth of $R(t)$ with $t$, i.e. the optimization has led the polymer to go through much deeper sites.}
\label{sitesalongpoly}
\end{figure}	

It is known that, in one dimension, the search for an optimal path in a disorder landscape leads to excursions larger than the thermal ones, which are of order $\sqrt{t}$. The shape of the optimal path strongly depends on the underlying disorder landscape, as shown in Fig. \ref{examplespoly}. In particular, the excursions are more important for a heavy-tailed disorder than for a Gaussian disorder. 
Those differences correspond to different optimisation strategies, as illustrated in Fig. \ref{sitesalongpoly}:
\begin{itemize}
\item
For a Gaussian disorder, the optimisation strategy is {\em collective}: the total energy of the polymer is equally shared between all the sites.
\item 
For an heavy-tailed pdf with $1< \mu <5$, the optimisation strategy is {\em elitist}: an important fraction of the total energy is hold by a small fraction of the sites belonging to the path.
\item
For an heavy-tailed pdf with $\mu<1$, the optimisation strategy is {\em individual}: most of the total energy of the polymer is localized on one particularly deep site.
\end{itemize}

Such differences in optimization have marked effects on the fluctuations properties at large $t$, in particular on the observables:
\begin{align}
&\overline {x(t)^2} \sim t^{2 \zeta} \\
& \overline{E(t)^2}^c= \overline{E^2(t)} - \overline{E(t)}^2 \sim t^{2\theta}\,.
\label{thetaandzeta}
\end{align}
Here $\theta$ and $\zeta$ are respectively the energy and the roughness exponents and show some universal features: they only depend on the behaviour of the disorder tails, namely the index $\mu$. Note that other important quantities, as the average energy $\overline{E(t)}$, strongly depend on all the microscopic details of the chosen model.

For a fast decaying disorder, the value of the exponents is known to be $\zeta = 2/3$ and $\theta = 1/3$ (\cite{huse_huse_1985,kardar_scaling_1987}) and has been recently proved, via mathematical (\cite{johansson_shape,prahofer_universal_2000}) and physical (\cite{Sasamoto_Spohn,calabrese_free-energy_2010}) approaches, for specific  fast decaying distributions such  as the Gaussian, the exponential or the LogGamma distribution
\cite{thiery2014log}.  For heavy-tailed disorder, where extremes play a major role in the choice of the optimal path, the values of the exponents appear to rely on the balance between the energy of the deepest sites and the deformation energy it would cost to reach them (\cite{zhang_growth_1990,krug_kinetic_1991,biroli_top_2007}).
Noting $t$ the length and $x$ the size of a typical excursion of the polymer, a result from extreme statistics of heavy-tailed distributions gives, for the volume $xt$ available to the polymer, an estimation of the energy of the deepest sites: $E_{min} \sim (xt)^{1/\mu}$. On the other hand, for the model with hard constraint the deformation cost is entropic and, provided that $x\ll t$, follows a scaling similar to an elastic energy as $S \sim -x^2/t$. Balancing both estimations, $E_{min} \sim S$, leads to the estimates:
\begin{align}
\zeta _{\mu} &= \frac{1 + \mu}{2 \mu- 1} \\
\theta _{\mu} &= \frac{3}{2 \mu - 1 } 
\end{align}

We can guess that those formula are valid for $2 \le \mu \le 5$, i.e. whenever $\zeta_{\mu} > 2/3$. Note that the values of the exponents are compatible with the scaling relation $\theta = 2 \zeta -1$. This relation comes from the statistical tilt symmetry (STS), originating in the invariance of the problem upon tilting transformation $x(\tau) \rightarrow x(\tau) + \epsilon \tau$ in the large scale limit \cite{schulz_brezin}.

Above $\mu=5$, the Flory argument leads to $\zeta_{\mu} < 2/3$: instead of a strategy focusing on deep sites of the disorder, the behaviour of the polymer is dominated by a {\em collective} strategy similar to the Gaussian case. 
On the other hand for $\mu=2$ we have $\zeta_{\mu} =1$ so that $x\sim t$ and the entropy can not be approximated by an elastic energy. Due to the hard constraint, the excursions of the polymer are confined in a cone. This observation leads to a new estimation of  the exponents $\zeta_{\mu} = 1$ and $\theta_{\mu}= 2/\mu $ for $0<\mu<2$. Note that
the STS symmetry is violated in that regime. 
All these Flory estimates are summarized in Table \ref{tableesti}. 
\begingroup
\renewcommand*{\arraystretch}{1.8}
\begin{table}
\centering
\begin{tabular}{|p{5mm}|c|c|c|c|}
  \hline
   & $\mu > 5$ & $5>\mu>2$ & $2>\mu>0$ \\
  \hline 
  $\theta_{\mu}$ & $1/3$ & $3/(2\mu -1)$ & $2/\mu$  \\
   \hline 
  $\zeta_{\mu}$ & $2/3$ & $(1+\mu)/(2\mu - 1)$ & $1$ \\
 
  \hline

\end{tabular}
\caption{ $\theta_{\mu}$ and $\zeta_{\mu}$  as a function of $\mu$. For $0<\mu<5$ the values of the exponents are estimated by scaling arguments. On the contrary in the {\em collective} optimization regime ($\mu >5$) no simple scaling argument is known.}
\label{tableesti}
\end{table}
\endgroup

When $\mu <1$, the first moment of the disorder distribution diverges. That leads to a huge separation of energy scale in the disorder, where all the sites can be neglected compared to the value of the most profound site through which the optimal path has to go.  Hence the optimization becomes {\em individual} and it allows to construct recursively the optimal path by picking the deepest site that pins the polymer, and applying the same strategy amongst the  sites inside the area delimited by the hard constraint.
 Such a hierarchical optimization strategy was coined {\em greedy} in \cite{hambly_heavy_2007}, where some of its properties were studied for the limit $\mu \rightarrow 0^+$. Here we will show that this approximation seemingly becomes asymptotically exact as $t \rightarrow \infty$ for all $\mu \le 1$. In particular the end point distribution of the polymer is computed analytically for the {\em greedy} strategy  and very well reproduces the numerical behavior for all $\mu<1$.

\section{Numerical simulations}

In this section, we present numerical simulations done with the matrix transfer method, which allows to keep track of both the energy and the position of the optimal path at every time $t$. 

We compare the numerical values of $\theta$ and $\zeta$ with the predictions of the (Flory) scaling arguments given in the the previous section, and obtain additional information about the whole pdf of the fluctuations of the total energy $E$.  The Flory estimates have been conjectured (see \cite{krug_kinetic_1991}) to be a good approximation only in the limit $\mu \rightarrow 2^+$. However a careful analysis of finite effects leads to the conclusion that the Flory argument may in fact be asymptotically exact \cite{PhysRevLett.69.3338}. Finally, we give strong numerical evidence of the existence of different optimization strategies as $\mu$ varies. This supports the correctness of the scaling argument in the regime of strong disorder for $\mu <5$.

\subsection{The scaling exponents}

\begin{figure}
\includegraphics[scale=0.25,trim=17mm 0 0 0]{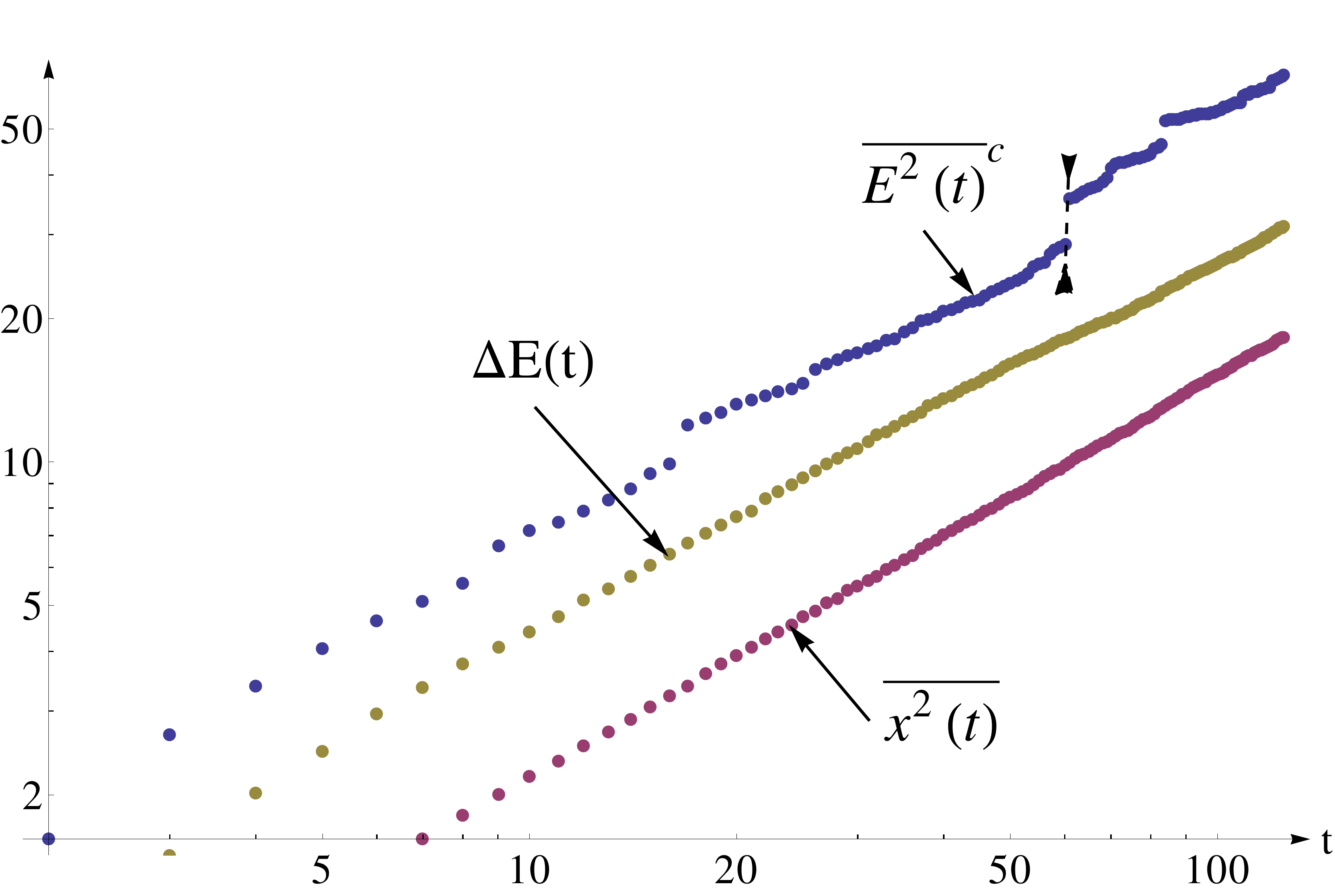}
\centering
\caption{Stability analysis of our numerical results. The averages are performed over $N=10^5$ samples. The mean squared displacement $\overline{x^2(t)}$ shows a well-defined smooth behaviour, while the variance of the energy $\overline{E^2(t)}^c$ displays a jump around $t=70$, which stems from a very deep single pin. This makes $\overline{E^2(t)}^c$ numerically unstable. On the contrary, for $\mu >2$, the quantity $\Delta E(t)$ displays a well-defined behaviour allowing for a reliable estimation of the exponent $\theta _ {num}$.}
\label{jumpestimators}
\end{figure}

To measure the exponents $\theta$ and $\zeta$, we can use the definitions given in Eq. \ref{thetaandzeta}. However, in Fig. \ref{jumpestimators}, we observe that the statistical estimator for $\overline{E^2}^c$ never averages out when $\mu<4$ and shows large jumps even for a very important sampling sizes. Note that the statistical estimator of $\overline{E^2}^c$ converges only if both $\overline{E^2}^c$ and its statistical error $(\overline{E^4}^c/N)^{1/2}$ are finite. But, due to the presence of heavy tails in the disorder, high enough moments of the distribution of energy $P(E)$ could diverge. We will see in Sec.\ref{theEpdf} that for $2<\mu<4$, $\overline{E^2}^c$ is finite while $\overline{E^4}^c$ diverges.

Another estimator of the spread of the distribution is the mean absolute deviation (MAD) $\Delta E$:
\begin{align}
\Delta E = \frac{1}{N}\sum_i |E_i - \overline{E}|
\end{align}
This estimator is more resilient to extreme events and works better with heavy-tailed distributions. Contrary to the standard deviation, which squares the distance from the average, MAD is well controlled as soon as the second moment of the pdf exists (in our case for $\mu>2$), and allows us to properly extract $\theta_{num}$ (see Fig. \ref{jumpestimators}). Note that $\overline{x^2(t)}$ does not present this kind of problem, because it is compactly supported due to the hard constraint (see Fig. \ref{jumpestimators}).

\begin{figure}
\includegraphics[scale=0.22]{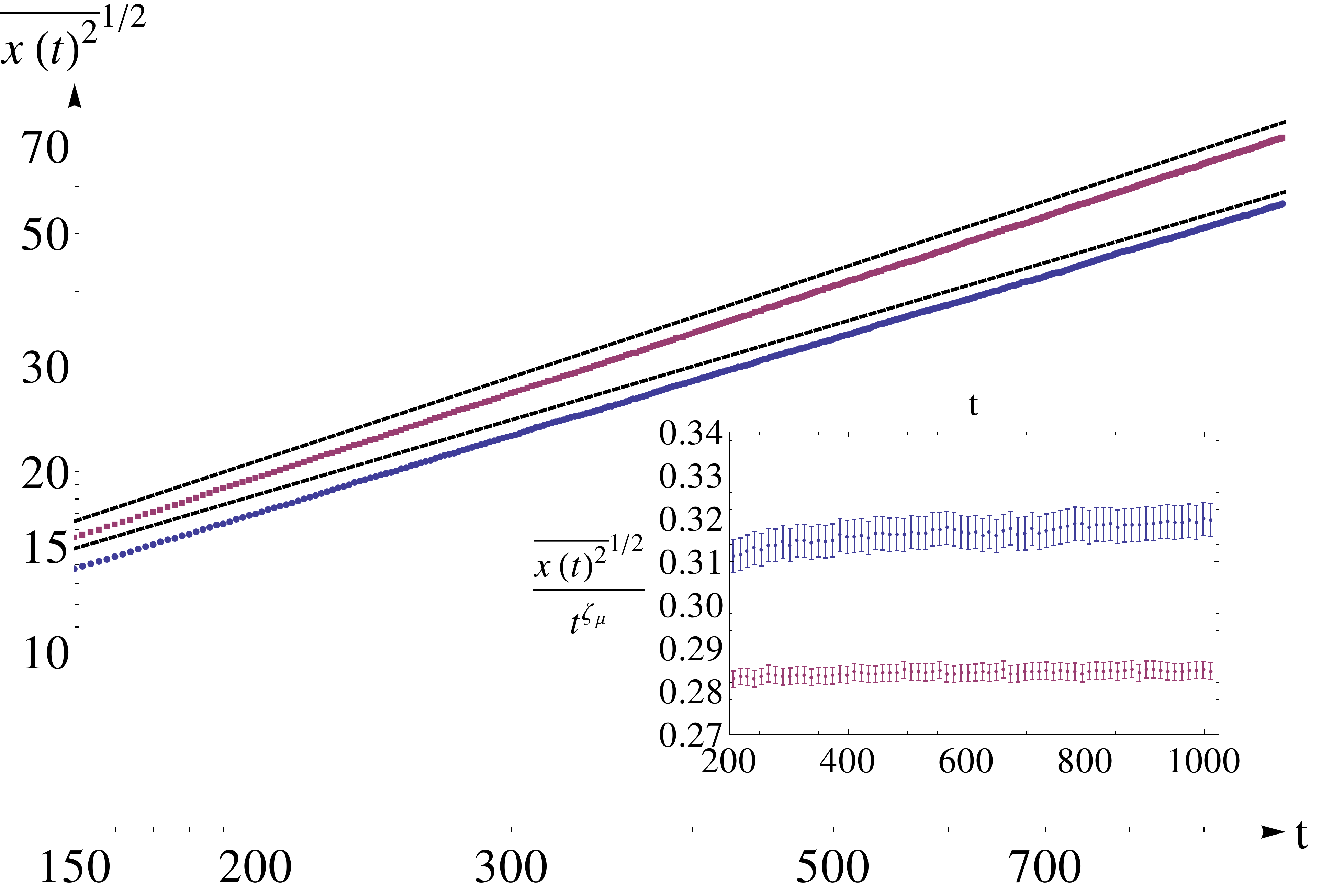}
\centering
\caption{Mean square displacement of the end position of the optimal path for $\mu=3$ (in red) and $\mu=4$ (in blue). Dashed lines correspond to the Flory estimate given in Table.\ref{tableesti}. Inset: $\overline{x(t)^2}/ t^{\zeta _{\mu}}$ showing saturation at large $t$ in both cases.}
\label{zetaest}
\end{figure}

\begin{figure}
\includegraphics[scale=0.23]{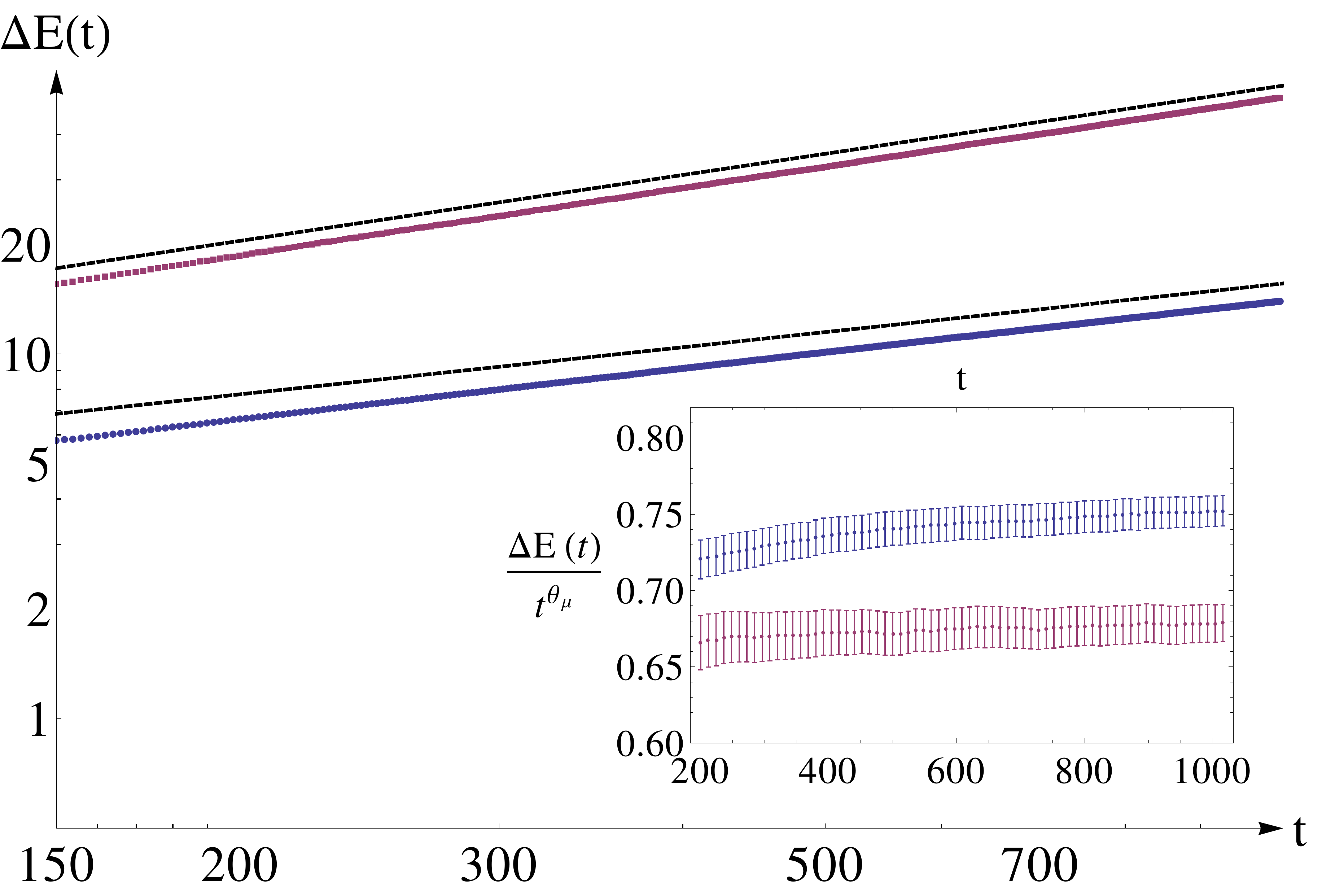}
\centering
\caption{Mean absolute deviation $\Delta E$ of the optimal energy for $\mu=3$ (in red) and $\mu=4$ (in blue). Dashed lines correspond to the Flory estimate given in Table.\ref{tableesti}. Inset: $\Delta E(t) / t^{\theta _{\mu}}$ showing saturation at large $t$ in both cases.}
\label{thetaest}
\end{figure}

The numerical values of the exponents for different values of $\mu$, and its comparison with the Flory prediction, are summarized in Table \ref{expotable}. The numerical estimations have been made with the maximum likelihood method. Figs. \ref{zetaest} and \ref{thetaest} illustrate 
how good the Flory prediction is: see in particular the insets where we show the quantities $\overline{x(t)^2}/ t^{\zeta _{\mu}}$ and $\overline{|E-\overline{E}|}(t) / t^{\theta _{\mu}}$ that should saturate to a constant when $t \gg 1$ if the scaling argument is correct. 
The saturation time should however grow larger and larger as $\mu \rightarrow 5^-$, explaining the observed difference between $\mu=3$ and $\mu=4$. Indeed, in this limit, the strategy remains {\em elitist}, but the effect of deep sites is not as strong and needs a large value of $t$ to be clearly distinguished from the rest of the ``crowd''. For $\mu>5$, the strategy becomes {\em collective}, and the exponents $\theta=1/3$ and $\zeta =2/3$ are recovered.

\begin{table}
\begin{center}
\begin{tabular}{|c|c|c|c|c|}
  \hline
  $\mu$ & $\theta _{\mu}$ & $\theta _{num}$ & $\zeta _{\mu}$ & $\zeta _{num}$ \\
  \hline
  3 & $3/5 =0.60$ & $0.605 \pm 0.006$ & $4/5 = 0.80$ & $0.802 \pm 0.004$ \\
  4 & $3/7 \simeq 0.43$ & $0.44 \pm 0.02$ & $5/7 \simeq 0.714$ & $0.715 \pm 0.005$ \\
  5 & 1/3 & $0.36 \pm 0.03$ & 2/3 & $0.69 \pm 0.04$ \\
  7 & 1/3 & $0.338 \pm 0.008$ & 2/3 & $0.669 \pm 0.004$ \\
  \hline

\end{tabular}
\caption{\label{expotable}Flory predictions of $\theta$ and $\zeta$ compared to numerical estimates for several values of $\mu$. The agreement 
is extremely good (see also Figs. \ref{zetaest} and \ref{thetaest}), except close to the transition value $\mu =5$, where the numerical estimate is less precise due to important size effects.}
\end{center}
\end{table}

\subsection{\label{theEpdf}Space and time fluctuations of the total energy}

In this section, we present results for the probability distribution of the fluctuations of the ground state energy $E$. The Gaussian (or fast decaying disorder) case has been studied extensively in the past. On the other hand, there is currenly no study of its equivalent in the heavy-tailed disorder.

\begin{figure}
\includegraphics[scale=0.31]{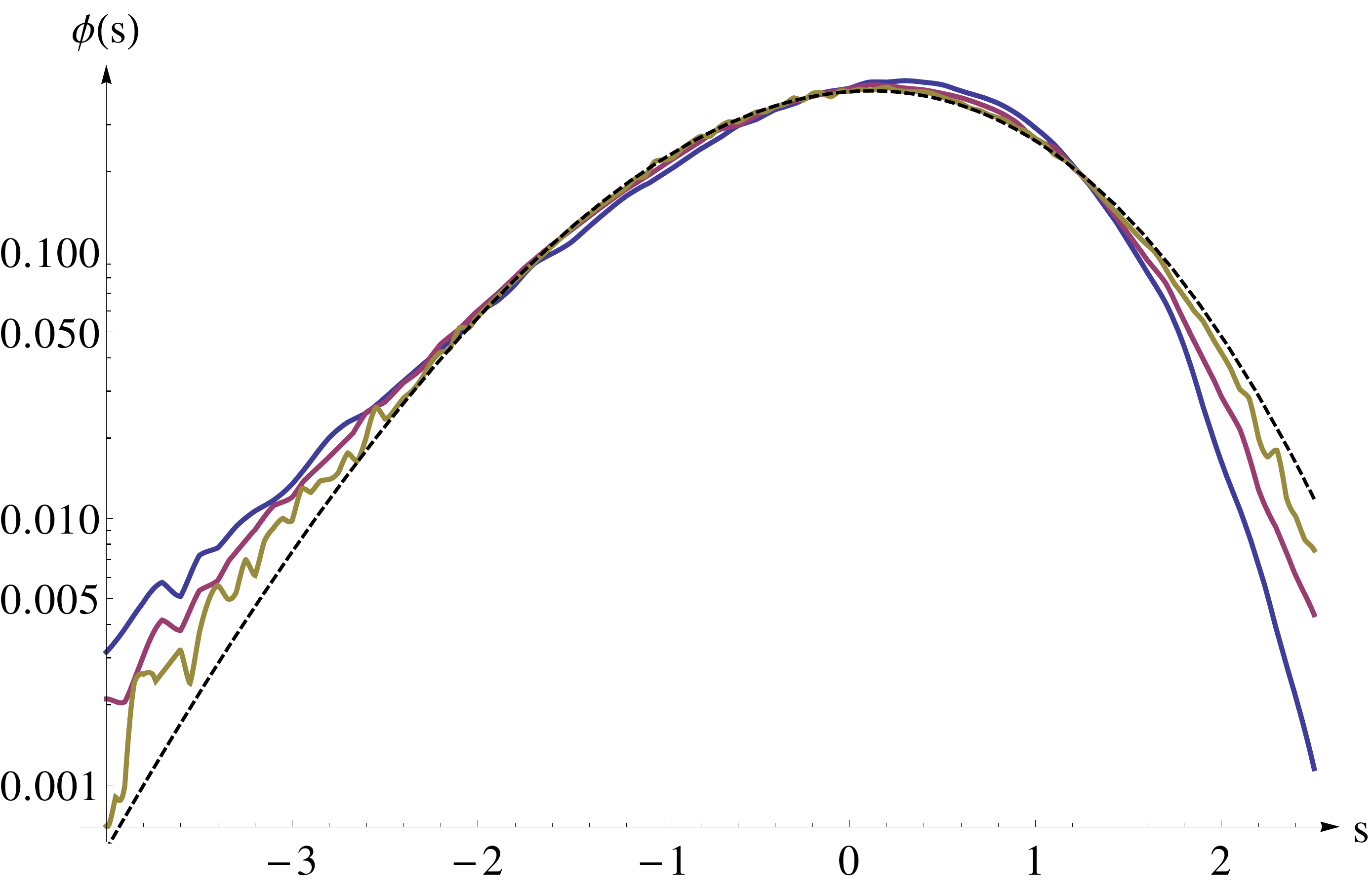}
\centering
\caption{Collapse of the pdf $\phi(s)$, for several lengths (from the farthest to the closest to the dashed line) $t=2^4 \text{ (blue)}, 2^7 \text{ (red)},2^8 \text{ (yellow)}$ for an disorder pdf with $\mu=8$. Comparison is made with the Tracy Widom distribution $F_2$ after centring and rescaling (in dotted black). Average over $N=2 \times 10^5$ samples.}
\label{mu8Edist}
\end{figure}

It has been shown that for some particular, fast decaying disorder distributions (see \cite{johansson_shape,calabrese_free-energy_2010,prahofer_universal_2000}), after the proper rescaling, the probability distribution converges to a family of distributions, called the Tracy-Widom (TW) distributions. It is believed that this universality extends to all fast decaying distributions. We define the rescaled variable:
\begin{align}
s(t)= (E(t) -\overline{E}(t))/(\overline{E^2}^c (t))^{1/2}
\end{align}
and compute in Fig. \ref{mu8Edist} the rescaled energy distribution $\phi(s)$ for different $\mu$'s. It seems numerically clear that the TW universality class extends for any disorder with $\mu > 5$.  Note that, for very negative $s$, $\phi(s)$ remains algebraic beyond some time dependent threshold $s< s^* _t$. However, when $t\rightarrow \infty$, the crossover towards the algebraic behaviour $s^* _t$ moves to $-\infty$.

\begin{figure}
\includegraphics[scale=0.43]{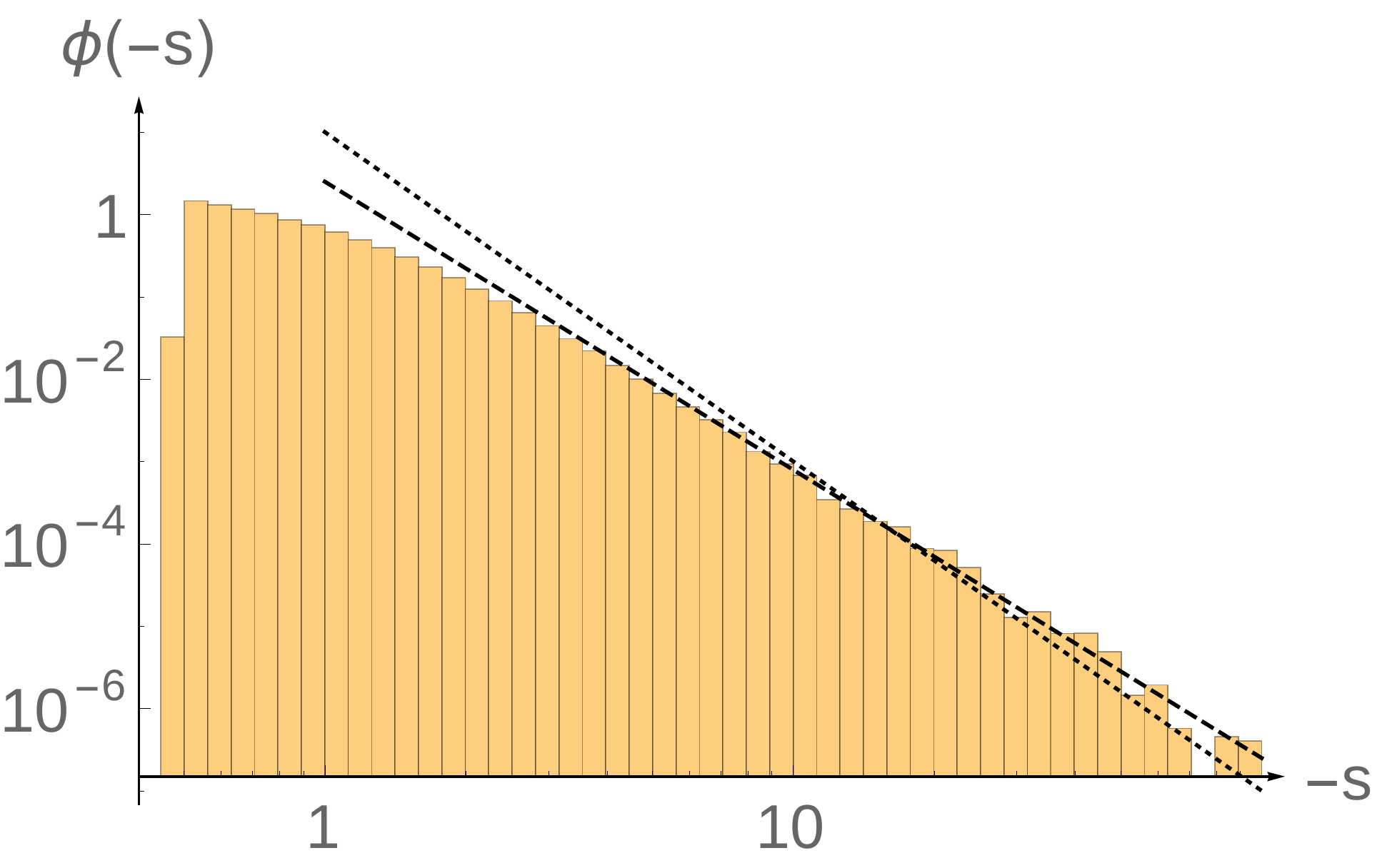}
\centering
\caption{The algebraic left tail of $\phi(s)$ on a log-log scale, compared with both decays of exponent $\mu$ and $\mu - 1/2$, for $\mu=3$. Polymers are of length $t= 2^{13}$, for a sampling of $N= 2 \cdot 10 ^6$ samples.}
\label{tail_prob}
\end{figure}

On the contrary, for $0<\mu <5$, the limiting distribution is very different. 
The family of limiting distributions $F_{\mu}$ depends only on $\mu$ and on the boundary conditions. Its analytical expression is still unknown; inspired by results from extreme statistics, a natural guess would be the Frechet distribution $\mathbb{P}(X<x)=\exp(- a |x|^{-\mu^\prime})$ or some convolutions thereof (see \cite{biroli_top_2007}). Similarly, one would intuitively assume that the tails exponents are preserved, as for the pdf of the minimum of independent random variables and so $\alpha = \mu$. However, the interplay with the elastic energy might lead to a shift in this exponent. This fact was already noticed in \cite{PhysRevE.89.042111} for the model of a particle in a disorder landscape confined by an harmonic potential. In this simpler model, the shift can be understood heuristically as follows: assuming again that the polymer is controlled by a very deep site, from record statistics, it is known that the tail of the minimum of $N$ heavy-tailed random variables decays as $\sim \frac{N}{|V|^{1+\mu}}$. Balancing elastic energy and potential leads to $E_{tot} \sim u^{2} \sim |V|$, and then to $N \sim u \sim |V|^{1/2}$. Hence the dependence of $N$ with $V$, inherent to the fact that large deviations in the disorder allows the particle to explore a larger space, which in turn leads to a modified exponent $\mu^\prime=\mu - 1/2$ of the tail of the total energy. Since for the heavy-tailed directed polymer case, the ground state is supposedly controlled by few deep sites, on which elastic energy and potential compete in a similar way, the above argument could also possibly describe the tails, and yield $\mu^\prime = \mu - 1/2$. However, this exponent is rather hard to extract from numerics, as it comes from large deviations in the free energy fluctuations, and hence from rare disorder realizations. Simulations up to $t= 2^{13}$ for more that $N= 2 \cdot 10 ^6$ samples nonetheless indicate a clear departure of $\mu^\prime$ from the value $\mu$ and tend 
indeed to favor the above conjecture $\mu^\prime = \mu - 1/2$ (see Fig.\ref{tail_prob}).

Let us add that those numerical results indicate that the support of the limiting distributions $F_{\mu}$ is the whole real line, unlike the standard Frechet distribution for heavy-tailed extreme statistics. An interesting feature of $F_{\mu}$ is the fact that its right tail, corresponding this time to unfavourable configurations of the disorder, appears to decay as $e^{- \alpha s^3}$, similarly to TW. This fact would support a mixture of some Frechet and TW distribution as a possible guess.

\begin{figure}
\includegraphics[scale=0.25]{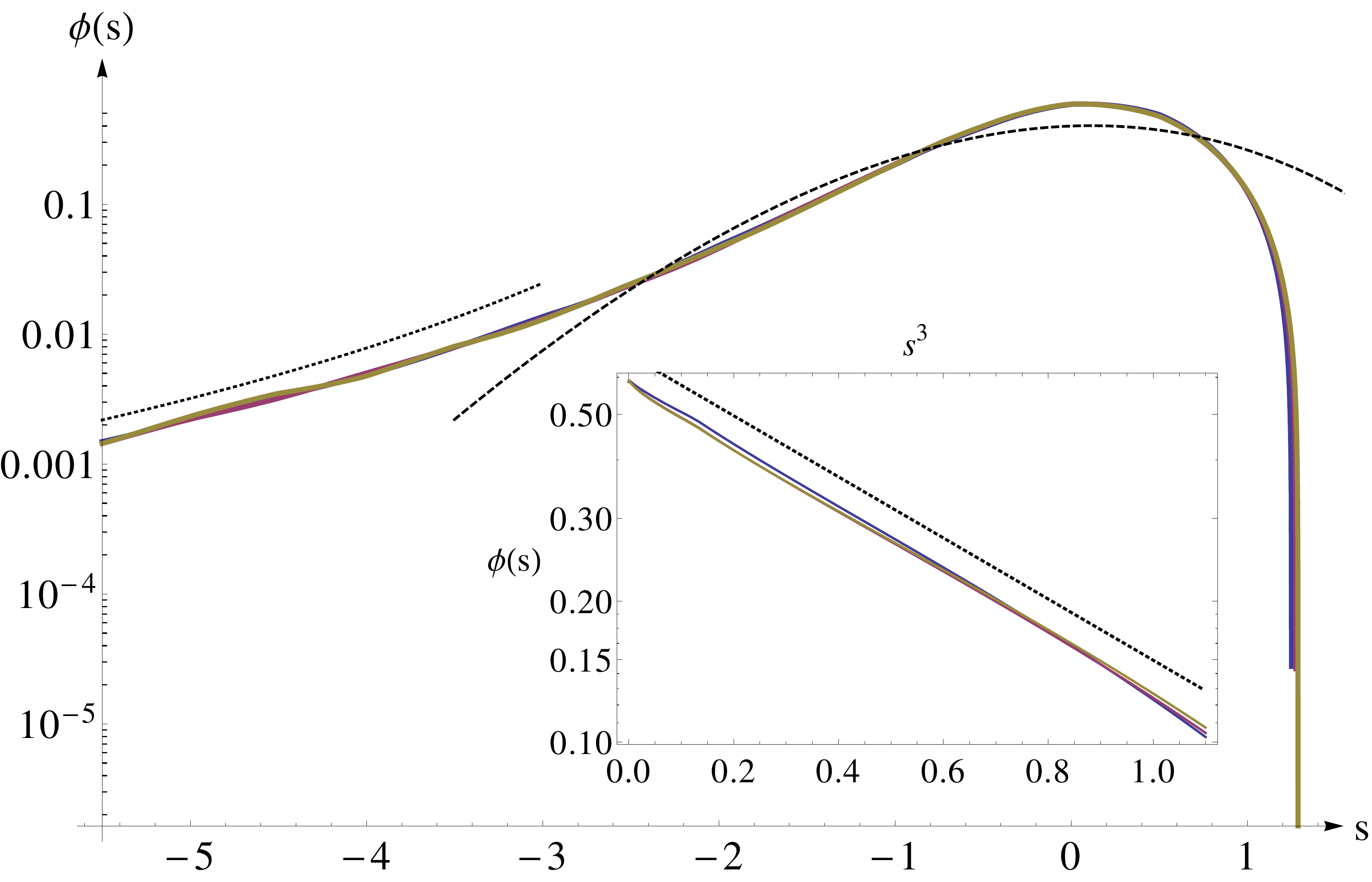}
\centering
\caption{Collapse of $\phi(s)$, for several lengths $t=2^4 \text{ (blue)}, 2^7 \text{ (red)},2^{10} \text{ (yellow)}$ and for an disorder pdf with $\mu=3$. Comparison is made with the Tracy Widom distribution $F_2$ after centring and rescaling (in dotted black). The decay estimate of index $\alpha = \mu \dfrac{•}{•}- 1/2$ is plotted in dashed black. Inset: far right tail of $\phi(s)$ compared to a decay of $e^{-1.5 s^3}$ (in dotted black). Average over $N=2 \times 10^5$ samples.}
\label{mu3Edist}
\end{figure}

For future reference, the tail analysis of $F_\mu(s)$ for the special case $\mu=3$ leads to the following numerical estimations: the left tail decays as $\int ^s dx \, F_3 (x) \approx 1.5 \, |s|^{-2.5}$ for $s \rightarrow - \infty$ while for the right tail $F_3 (s) \approx 1.8 \, e^{-1.5 s^3}$ for $s \rightarrow \infty$ (see Fig. \ref{mu3Edist}).

\begin{figure}
\includegraphics[scale=0.31]{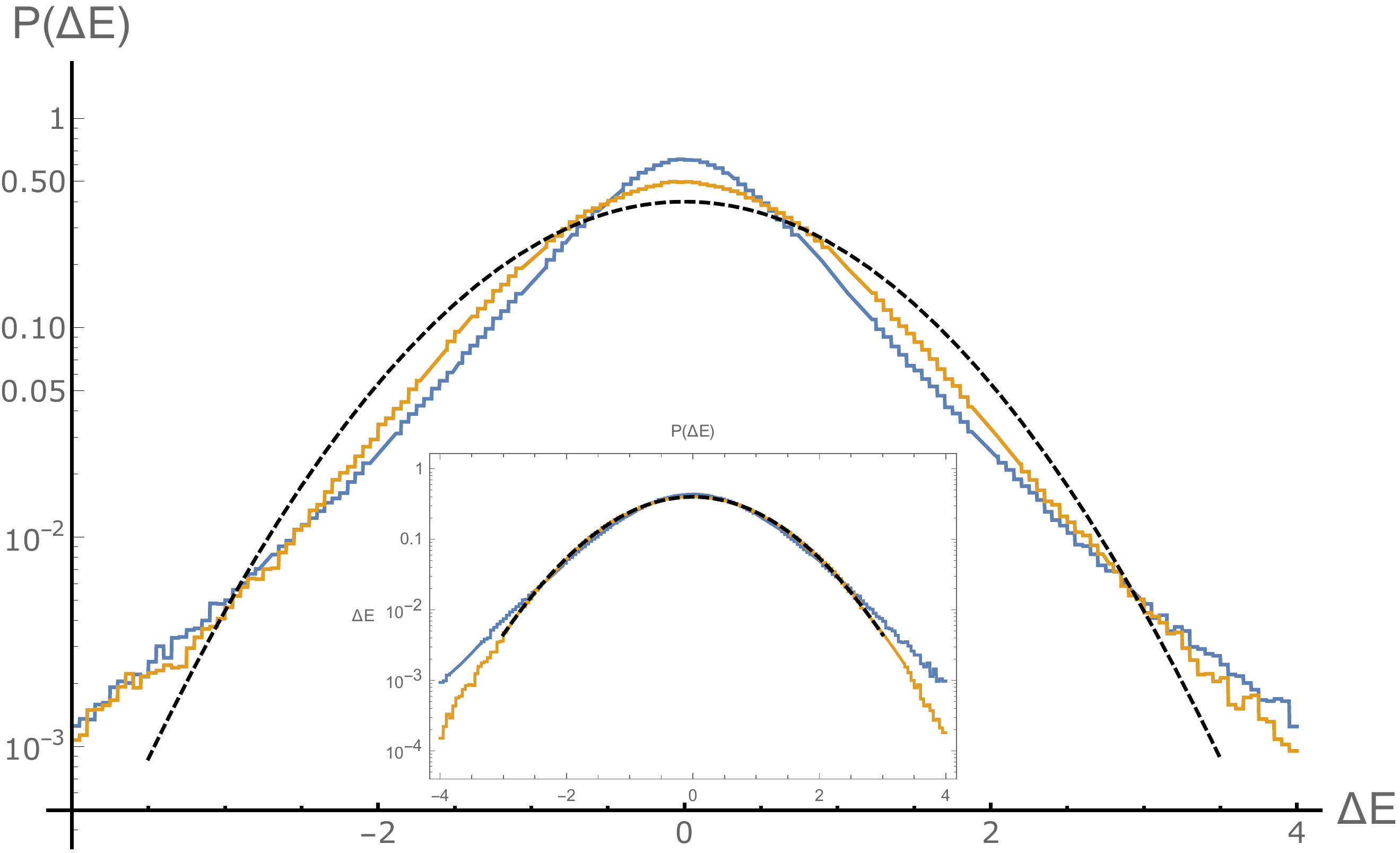}
\centering
\caption{Probability distribution of $\Delta E$ for various sizes of the increments $\Delta x$ of the stationary distribution for the disorder with $\mu=3$: from the most peaked to the least, $\Delta x=4 \text{ (blue) and } 64 \text{ (yellow)}$. The variance of the distributions is normalized to $1$. Tails survive even for large increments. The dashed black line is the Gaussian distribution. Inset: same probability distributions for a disorder with $\mu=8$, with convergence towards a gaussian distribution at large increments. The time used in simulation is $t=10^5$.}
\label{gaussconv}
\end{figure}

{
Let us now turn to the dependence of the ground state energy on its final position, $E(x,t)$, when the latter is imposed. In the Gaussian case, for infinite systems at large time, this process has been characterized as an Airy process, dependent of the initial conditions. For small $x$, $E(x,t)$ behaves as a mere Brownian motion in $x$, but its correlations saturate for $x \gg t^{2/3}$. [It is worthwhile to mention that, in the discrete grid model used presently, $E(x,t)$ does not have Gaussian spatial increments, and these increments are in fact weakly correlated. Nonetheless, the correlation length does not grow with $t$ and in the continuum limit  $E(x,t)$ indeed converges (for small $x$) to a Brownian motion]. For a heavy-tailed disorder with $\mu >5$, our results are consistent with the scenario where this convergence
holds, as can be seen in the inset of Fig. \ref{gaussconv}.

The situation is quite different in the \textit{elitist} optimization strategy. Because the increments of the stationary  process are controlled by the deep sites, they exhibit jumps of all sizes with power-law tails. Furthermore, the increments are, through scaling arguments, expected to behave as $\Delta E (x) \sim x^{\theta/\zeta}$ with $\theta/\zeta = 3/(1+\mu) > 1/2$ as soon as $\mu <5$. Hence the time stationary process of $E(x,t)$ is super-diffusive (in $x$), with strongly correlated heavy-tailed increments of index $\mu$ (see Fig.\ref{gaussconv}). A more precise characterization of this process (which is neither a fractional Brownian motion nor a L\'evy walk) would be quite interesting and is left for future investigations. 

}

\subsection{Zooming into the optimisation strategies}

Although the scaling argument gives the correct estimates, it relies on the assumption that the fluctuations of $E$ are controlled by the fluctuations of the deepest sites in the disorder. This stems from the fact that the optimal path does large excursions specifically to reach some favorable pinning sites.
Note that, at variance with the case $\mu<1$ where one site dominates over all the the others, when $1<\mu< 5$ there is still a large population of sites 
with local energy of order $V_{min}$ -- there is in that case no obvious dominance. Hence to check the validity of the {\em elitist} optimisation strategy, one has to test the fact that the optimal path indeed picks up {\it some} sites amongst the deepest available, whereas in the {\em collective} case this is not so. 

We have therefore formulated the {\em elitist} optimisation hypothesis as follows: Consider a paraboloid region ${\cal R}(\alpha)$ of length $t$ and width $t^{\alpha}/2$, containing a number of sites $\propto t^{1+\alpha}$. We then compute the probability $P_c(\alpha)$ that the minimum energy site along the polymer is one among the $\mu \log(t)$ deepest sites within the region ${\cal R}(\alpha)$. [The choice of $\log t$ sites corresponds to a rough estimate of the number of completely independent paths in region ${\cal R}(\alpha)$.] 

Intuitively, we expect that for the {\em elitist} strategy, the minimum energy along the polymer should be deeper than the minimum restriced to ${\cal R}(\alpha) \subset {\cal R}(\zeta)$ when $\alpha < \zeta$. On the contrary, whenever $\alpha > \zeta$, the minimum along the polymer should be higher than the minimum in ${\cal R}(\alpha)$, since the elastic energy prevents it from reaching this particularly favourable site. Therefore, when $t \rightarrow \infty$, $P_c(\alpha)$ should become more and more peaked around $\alpha = \zeta$ for the {\em elitist} strategy. In the {\em collective} regime, on the other hand, $P_c(\alpha)$ should vanish for all $\alpha$ since the global optimisation has nothing to do whatsoever with the value of the deepest 
available site. 

\begin{figure}
\includegraphics[scale=0.24]{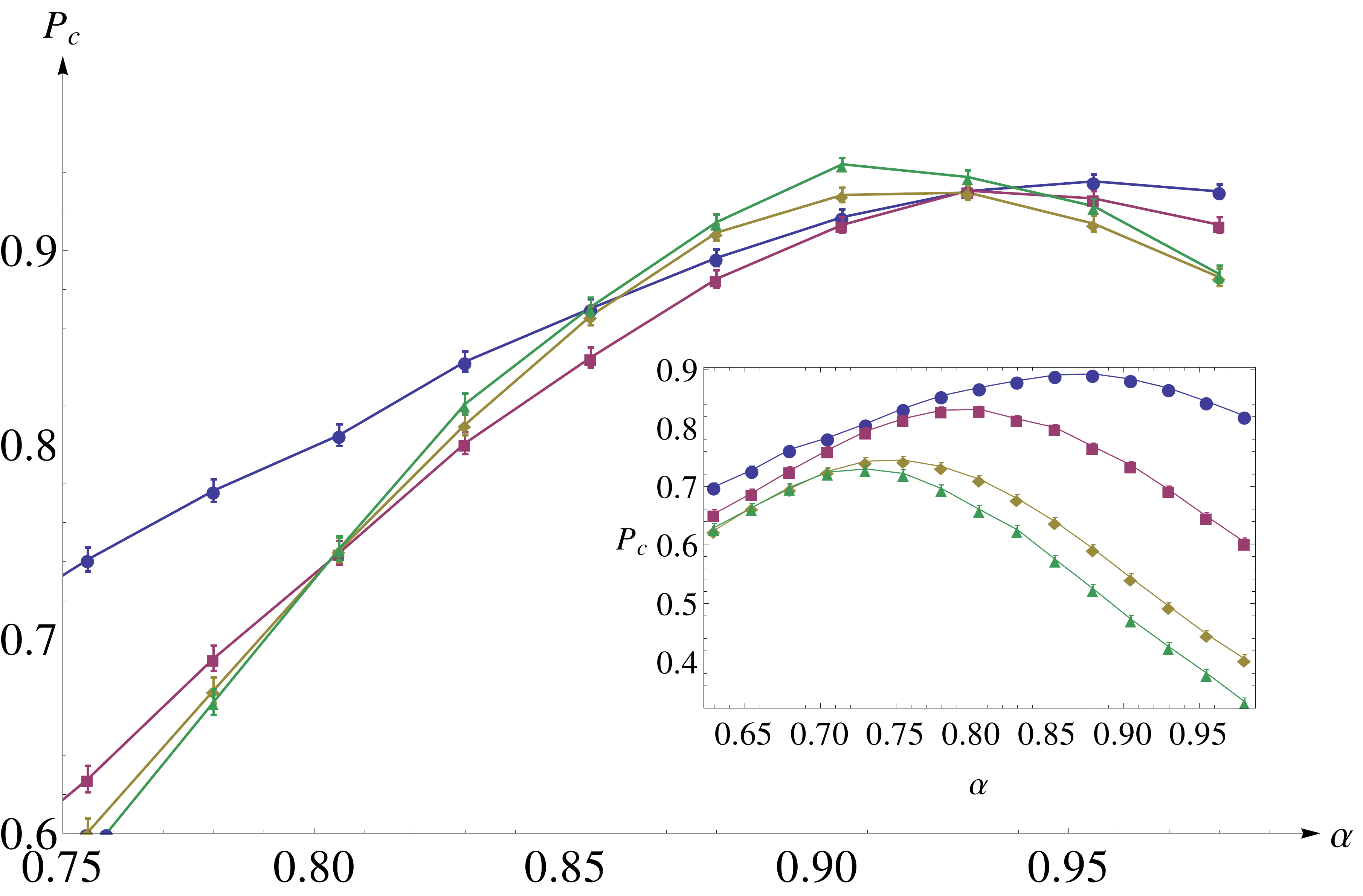}
\centering
\caption{Probability $P_c(\alpha)$ as defined in the text for a heavy tailed disorder with $\mu =3$, corresponding to an {\em elitist} optimisation. Symbols correspond to different lenghts: $t=2^6 \text{ (circles)},2^9\text{ (squares)},2^{11}\text{ (losanges)},2^{12}\text{ (triangles)}$. The averages are performed over $N=2 \times 10^5$ samples. As expected, $P_c(\alpha)$ appears to peak around the value $0.9$, which should converge to $\zeta_3 = 0.8$
for very large sizes. Inset: the same analysis performed for $\mu=8$, corresponding to a collective optimisation. In this case, $P_c(\alpha)$ decreases 
as a function of $t$, indicating that the minimum energy site becomes less and less relevant as the size of the polymer increases.}
\label{picrugo}
\end{figure}

\begin{figure}
\includegraphics[scale=0.32]{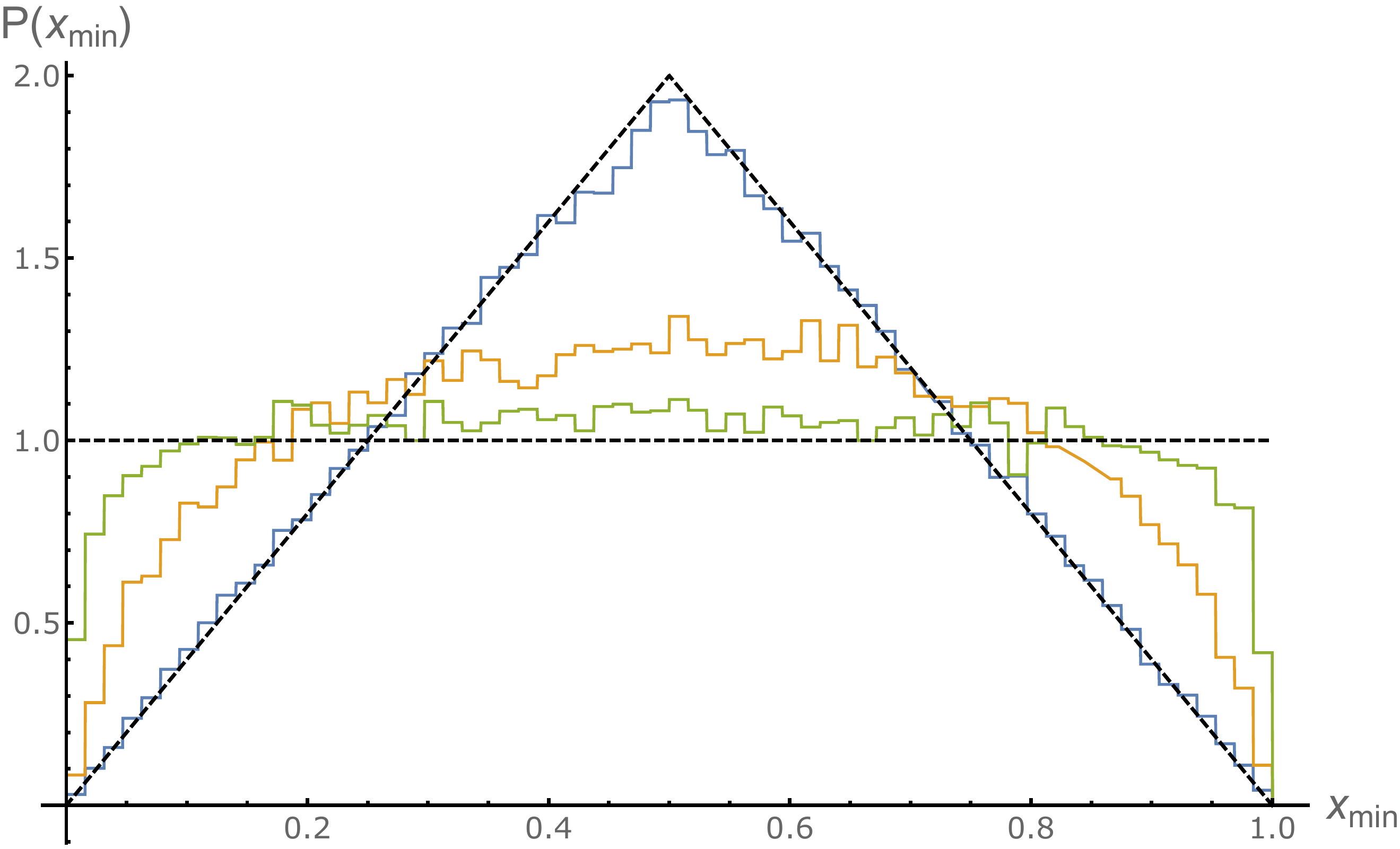}
\centering
\caption{Histograms for $P(x_{min})$ of the position of the minimum $x_{min}$ (normalized in $[0,1]$) of site energy \textit{along} the polymer in the with both extremities fixed, for various disorder. From the most peaked to the flattest, $\mu = 0.5 \text{ (blue) },  3 \text{ (yellow) }$ and Gaussian disorder (green). In dashed black, both the uniform distribution and the distribution of the minimum for the greedy case, to which disorders with $\mu <1$ converge. $t=512$ and $N=10^5$ samples. We expect that the distribution converges to a uniform distribution at large $t$ for $\mu > 5$, but to remain non trivial for $\mu < 5$.}
\label{posimin}
\end{figure}

Fig. \ref{picrugo} suggests a different qualitative behaviour for $\mu=3$ and $\mu=7$, in agreement with the above prediction. 
Note that the maxima of the curves in Fig. \ref{picrugo} corresponding to an estimation of the rugosity exponent $\zeta$ are moving to the left as $t$ increases. They are expected to converge towards $\zeta_{3} = 4/5$ for $t \to \infty$, although the convergence appears to be very slow.
Such a difference in optimization strategy is reflected in the extremal statistics observed \textit{along} the polymer path. Granted the Gaussian optimal path is not relying on extremal sites, the position of minimum energy sity should be (asymptotically) flat over $[1,t]$ (see Fig. \ref{posimin}), with finite size effects coming from the smaller entropy at the very extremities of the polymer. On the other hand, for $\mu < 5$ and for periodic boudary conditions, the position of the minimum site is related to the maximum entropy of the polymer, hence the mode should be located right in the middle of the path. In the greedy case $\mu \rightarrow 0^+$, the probability distribution of this minimum is easily computed and represented in Fig.\ref{posimin}, as the limiting distribution for every disorder with $\mu <1$.

\section{Distribution of the end point}

Another observable of interest is the distribution of the end point of the (unconstrained) optimal path. Compared to the fluctuations of the energy $E(t)$, we know much less about the statistics of $x(t)$. One expects the rescaled position $z= x(t)/t^{\zeta}$ to converge in law towards a limiting distribution $Q_{\zeta}(z)$, but the analytic shape of $Q_{\zeta}(z)$ is unknown even in the Gaussian case $\zeta=2/3$. For an exponential distribution of the disorder, the whole process $E(t,x)$ is characterized as the so-called Airy process \cite{johansson_discrete_2003}, which allows one to extract the joint distribution of the position and total energy of the optimal path $\mathbb{P}(E(t),x(t))$. Although the marginal $Q_{2/3}(z)$ cannot be computed explicitly, it has been shown that it has an infinite support with a rather weak departure from the Gaussian distribution \cite{schehrvicious,flores_endpoint_2013}. 

The heavy-tailed disorder case exhibits a radically different behaviour, since $Q_\zeta(z)$ is strongly influenced by the large excursions of the optimal path to reach pinning sites. For $\mu <2$, $\zeta$ saturates to $1$ due to the hard constraint and the support of $Q_\mu(z)$  reduce to the interval $ [-1/2,1/2]$: the extremity has a finite probability to reach any point of the available space, even at large $t$.

\begin{figure}
\includegraphics[scale=0.22]{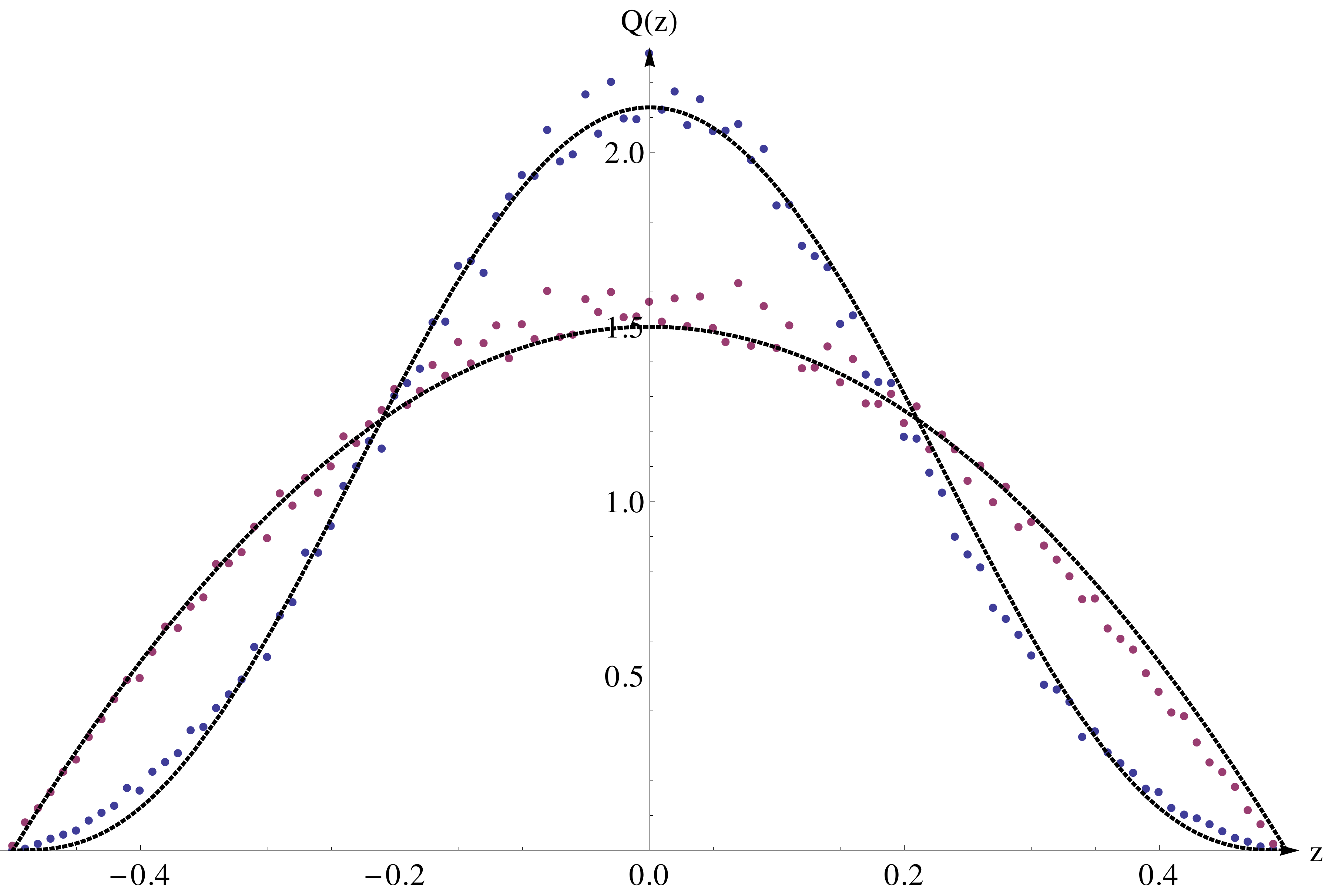}
\centering
\caption{The pdf $Q(z)$ of the position of the free end as a function of $z=x(t)/t$. In blue for a disorder pdf of $\mu=1.5$, in red $\mu=0.5$. In dotted black, the theoretical parabolic prediction. We also show the symmetric Beta distribution fit, with $\nu = 3.8$. $t=2^{11}$ and $N=2 \times 10^5$.}
\label{extremityP}
\end{figure}

Denoting $z = x(t)/t \in [-1/2,1/2]$, the distribution $Q_{\mu}(z)$ can be explicitly computed for the {\em greedy} strategy, where the optimization becomes a hierarchical recursive process. In Appendix \ref{appA} we give the details of the computation and our final result reads:
\begin{align}
\label{parabolaformula}
Q_{\text{greedy}}(z)=6 \left( \frac{1}{4}-z^2 \right)
\end{align}
This results becomes exact for very small $\mu$,  but, in Sec.\ref{model} it was argued  that the greedy strategy should asymptotically hold for every $\mu<1$. This assumption is further confirmed by numerics for $Q_\mu(z)$ (see Fig. \ref{extremityP}), retaining its parabolic shape until $\mu=1$. For $\mu >1$, the support still remains the interval $[-1/2,1/2]$, but $Q_{\mu}(z)$ is modified. The numerical results are relatively well fitted by the Beta distributions family $\mathbb{B}(\nu,\nu) = c_{\nu} (1/4 - x^2)^{\nu}$, where $\nu$ is a fitting parameter that depends on $\mu$ (see Fig.\ref{extremityP}).  

Let us now consider the case $2 < \mu < 5$. 
We could expect that the tails of $Q_{\mu}(z)$ are again controlled by the position of the deepest available site. The decay exponents may therefore be inferred from the zero-dimensional particle model, as in Section \ref{theEpdf} above. This approach predicts \cite{PhysRevE.89.042111}:
\bea \label{qmu} 
Q_{\mu} (z) \sim \frac{1}{z^{2 \mu}}.
\eea
In practice, it is delicate to estimate accurately the tail exponent of $Q_\mu(z)$, because of the presence of the cut-off imposed by the hard constraint of our discrete numerical model. Our results are nevertheless
consistent with the above prediction Eq. (\ref{qmu}). 
Such an algebraic tail is important conceptually when comparing with e.g. the Functional RG approach. 
Indeed moments of the position are related to the moments of the renormalized disorder in
that method \cite{LeDoussal201049}. In our case moments of sufficently high orders diverge,  implying that the loop expansion
of the standard Gaussian disorder case \cite{FRG_chauve} needs to be reconsidered as soon as $\mu < 5$, even when the two-body correlator
of the disorder is perfectly defined (i.e. when $\mu > 2$).

\section{Conclusion}

We presented a detailed investigation of the modification of the KPZ universality class due to heavy-tailed disorder, building upon several previous studies
\cite{zhang_growth_1990,krug_kinetic_1991,hambly_heavy_2007}. These modifications are deep and to some 
extent unexpected since the KPZ universality class breaks down as soon as the fifth moment of the disorder diverges -- when one would have naively expected that the convergence of the second moment (the variance) of the disorder would suffice. As we emphasized in the introduction, this paradox   becomes even more acute above the upper critical dimension -- see \cite{biroli_top_2007}. From a KPZ point of view, this shows that the 
interplay between non-linearities and rare but large events is highly non trivial and requires new analytical methods. Nonetheless, new universality classes emerge, since both the exponents and corresponding probability distributions seem uniquely controlled by the behaviour of the tail, the bulk of disorder distribution being irrelevant. 

However, our hope to get a better grasp on the KPZ class itself by studying its surrounding is clearly thwarted by the dearth of analytical tools adapted to heavy-tailed noise. Exact results are scarce and numerical simulations have been extensively used to enlighten the situation. They indicate a sharp transition when the fifth moment of the distribution diverges, between the Gaussian KPZ/DP regime 
characterized by the standard exponents $(\theta=1/3,\zeta=2/3)$ and some non-trivial values $(\theta_{\mu},\zeta _{\mu})$ that seem to be exactly predicted by a simple Flory-like argument. The full ground state energy distribution also changes from Tracy-Widom to a new family of functions, with a power-law tail for large negative values of the energy and the Tracy-Widom $e^{-s^3}$ behaviour for anomalously ``bad'' ground state energies. Several other quantites have been investigated -- such as the process governing the energy as a function of the end point of the polymer or the distribution of this end point -- which are also markedly different from their Gaussian counterpart in the tail-dominated regime. Finally, we have attempted to obtain direct
evidence that the deepest available sites play a special role in the tail-dominated regime, giving support to the validity of the Flory argument. Still, 
there are many outstanding open problems. For example, the largest eigenvalue of random matrices with heavy-tailed distributed entries are known to leave the Tracy-Widom universality class when the fourth moment (rather than the fifth) diverges \cite{biroli_top_2007}. This questions the existence of an exact mapping between the DP problem and random matrices ensembles for heavy-tailed disorder (see the discussion in \cite{biroli_top_2007}). 

A very important issue (in our minds) is the extension of the Functional RG approach to heavy-tailed situations. This might shed considerable light on the
method itself, and on the deep underlying mechanisms at play in pinning problems and in non-linear (KPZ like) stochastic partial differential equations. Our detailed numerical study of the directed polymer has unveiled that the new statistics appears to possess an underlying recursive structure. This 
hypothesis is backed by the solvable ``greedy'' limit of an exponentially wide disorder. This calls for the development of theoretical tools able to handle these hierarchical processes.

\appendix
\section{Appendix: Derivation of $Q_{\mu}(x)$ in $\mu \rightarrow 0$ \label{appA}}

The derivation is eased by taking the continuum limit, where for convenience we rescale the position of the end point to $z \in [-1,1]$. We introduce the sequence of variables $\xi_i =\frac{x_i + y_i}{2}$ and $2 r_i = y_i - x_i$, $(x_i,y_i)$ being the coordinates of the pinning site chosen at step $i$. The measure being uniform over the space of intervals, it stays uniform if we fix the center of mass, under the constrain that the end points cannot leave $[-1/2,1/2]$. The joint probability distribution is, constrained on $[-1,1] \times [0,1]$:
\begin{align}
P_0(\xi,r) d\xi dr = \Theta (r \leq 1 - |\xi|) d\xi dr
\end{align}

Due to self similarity of the process, there are recursive relations between the end point after $i$ steps and $i+1$ steps. We are eventually interested in the limit of the following process $z_{\infty}$, describing the position of the end point at $n\rightarrow \infty$:
\begin{align}
z_{\infty} = \xi_1 + r_1 \xi_2 + r_1 r_2 \xi_3 + \cdots
\end{align}

All couples $(\xi _i, r_i)$ having the same joint distribution $P_0$ and being independent for $i \neq j$. this bears some similarities with Kesten variables \cite{Kesten} but note that $\xi _i$ and $r_i$ are not independent themselves. The variable $z_{\infty}$ obeys the following equation:
\begin{align}
z_{\infty} =_{law} \xi + r z_{\infty}
\end{align}

This leads to an integral equation for $P(z_{\infty})$ the PDF of the end point, for example if we choose to condition over the value of $z_{\infty}$ in the above equation:
\begin{align}
\nonumber
P(z)&= \int _{r,u} P_0 (z- r u,r) P(u) du \, dr \\
\label{integraleqprob}
&= \int_{r,u} \Theta (r < 1 - |z -r u| ) P(u) du \, dr
\end{align}

Although there is no generic way to solve such integral equations, we can recursively compute the moments or use the above equation to write down a differential equation for $\phi (\lambda) = E(e^{i \lambda z _{\infty}})$. Or one can simply check that a parabola $P(z) = \frac{3}{4}(1-z^2) $ is the proper solution.
$\Theta (r < 1 - |z-r u|)$ is non zero for $r < \frac{1+z}{1+x}$ if $z<x$ and $r < \frac{1-z}{1-x}$ if $x<z$. Hence, the right side of Eq.\ref{integraleqprob} is equal to:
\begin{align}
\nonumber
\frac{3}{4} \left( \int _{-1 < x <z} \frac{1+z}{1+x} (1-x^2) dx \right. \\
\nonumber
\left. + \int _{z<x<1} \frac{1-z}{1-x}(1-x^2) dx \right) \\
\nonumber
= \frac{3}{4}(1-z^2)
\end{align}
And the result follows as given Eq.\ref{parabolaformula} after the rescaling $z \rightarrow z/2$.

\bibliography{hillyland}
\end{document}